\documentclass[letterpaper,10pt]{article} 

\usepackage{osameet3} 

\newcommand\authormark[1]{\textsuperscript{#1}}

\usepackage{amsmath,amssymb}
\usepackage[colorlinks=true,bookmarks=false,citecolor=blue,urlcolor=blue]{hyperref} 

\begin{document}

\title{Turbulence-free computational ghost imaging}


\author{Qiang Gao,\authormark{1} Yuge Li,\authormark{1} Yunjie Xia,\authormark{1,2} Deyang Duan \authormark{1,2,*}}

\address{\authormark{1} School of Physics and Physical Engineering, Qufu Normal University, Qufu 273165, China\\
\authormark{2}Shandong Provincial Key Laboratory of Laser Polarization and Information
Technology, Research Institute of Laser, Qufu Normal University, Qufu 273165, China}

\email{\authormark{*}duandy2015@qfnu.edu.cn} 

\begin{abstract}
Turbulence-free images cannot be produced by conventional computational ghost
imaging because
calculated light is not affected by the same atmospheric turbulence as
real light. In this article, we first addressed this issue
by measuring the photon number fluctuation autocorrelation of the signals
generated by a conventional computational ghost imaging device. Our results
illustrate how conventional computational ghost imaging without
structural changes can be used to produce turbulence-free images.
\end{abstract}

\section{Introduction}
Ghost imaging with thermal light, in its most basic form, is an indirect
imaging technique. Conventional ghost imaging requires two light beams: the
reference beam, which never illuminates the object and is measured directly
by a charged-coupled device (CCD), and the signal beam, which illuminates
the object and is measured by a detector with no spatial resolution (bucket
detector). The ghost image is restricted by a coincidence measurement of the
signals from the two detectors. However, dual optical ghost imaging is not
suitable for applications such as remote sensing [1,2], 
lidar sensing [3-5] and night vision [6-8]. Fortunately, Shapiro proposed computational ghost
imaging in 2008 [9]. In this technique, the CCD detector is
replaced with a virtual detector that calculates the propagation of the
field of the reference beam. The image is reconstructed by correlating the
calculated field patterns with the measured intensities at the object plane.
As a result, computational ghost imaging, similar to classical optical
imaging, only has one optical path, which has attracted increasing attention [1,6,7,10-13].

Although ghost imaging technology has many advantages, such as
superresolution [14] and indirect imaging capabilities [15,16], its most valuable property is its incredible
turbulence-free imaging capability. This practical property is an important
milestone for optical imaging because any fluctuation index disturbance
introduced in the optical path will not affect the image quality [17-19]. However, previous research has shown that
turbulence-free imaging has certain conditions that must be met [20,21]. One of the critical conditions is that two optical paths
need to pass through the same turbulence. For computational ghost imaging,
there is only one optical path. This optical path is affected by atmospheric
turbulence, while the other virtual optical path is obtained through
calculation. Thus, these two optical paths cannot be affected by the same
atmospheric turbulence. Thus, turbulence-free images cannot be obtained from
conventional computational ghost imaging [19]. Although some
methods, such as adaptive optics and depth learning, have significantly
improved the image quality of computational ghost imaging in turbulent
environments, these methods are difficult to produce a turbulence-free
image [22-24].

To solve this challenging problem, we used a conventional computational
ghost imaging device as a signal acquisition terminal. The signals produced
by this device were duplicated into two identical copies. Thus, both sets of
data were affected by the same atmospheric turbulence. As a result,
turbulence-free images were produced. Finally, the turbulence-free image was
restructured by measuring the photon number fluctuation autocorrelation of
the two signals.

\section{Theory}

\begin{figure}[h]
\centering\includegraphics[width=1\linewidth]{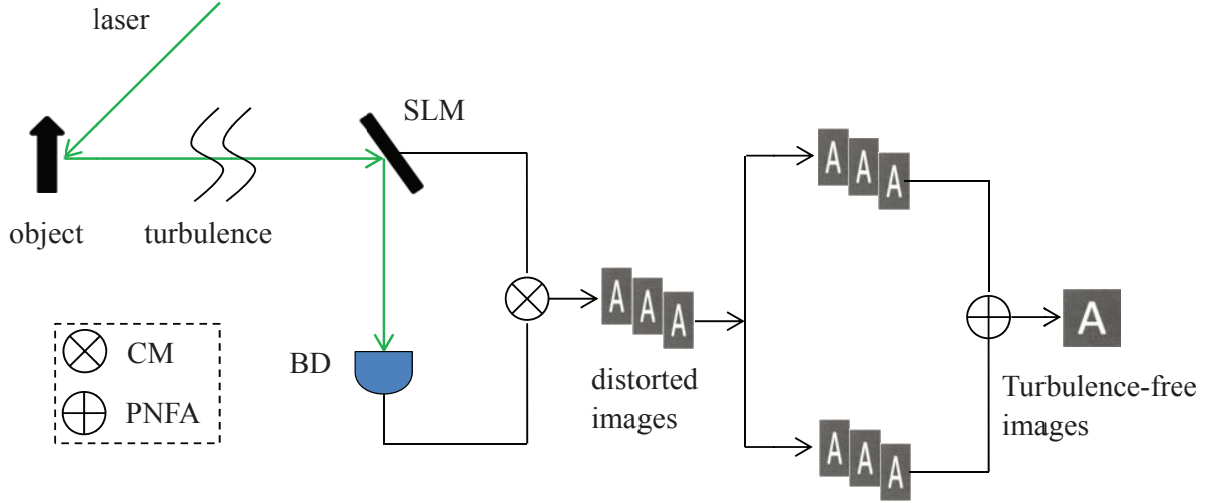}
\caption{(a) Setup for turbulence-free computational ghost imaging. SLM:
spatial light modulator, BD: bucket detector. CM: coincidence measurement,
PNFA: photon number fluctuation autocorrelation.}
\end{figure}

The setup for turbulence-free computational ghost imaging is illustrated in
Fig. 1. A laser beam illuminated an object $T(x)$; then, the reflected light
that passed through atmospheric turbulence was received and modulated by a
spatial light modulator (SLM). The light intensity was detected by a bucket
detector and can be expressed as $\sum_{p}E\left( x,t\right) T\left(
x\right) e^{i\phi }$, where $p$ represents $p$th subfield from the $p$th
subsource, $e^{i\phi }$ represents the atmospheric turbulence factor [17,18,22-26]. Because the
calculated light is based on the input signal of the SLM, it is not affected
by atmospheric turbulence [9,23,24,27]. Thus, the calculated
light can be expressed as $\sum_{q}E\left( x,t\right) $, where $q$ represents $q$th subfield from the $q$th
subsource. The signal produced
by the conventional computational ghost imaging device can be expressed as [9,17,21,24,25]
\begin{equation}
C\left( x,x^{^{\prime }}\right) =\left\langle \sum_{p}\left\vert E\left(
x,t\right) T\left( x\right) e^{i\phi }\right\vert ^{2}\sum_{q}\left\vert
E\left( x,t\right) \right\vert ^{2}\right\rangle -\left\langle
\sum_{p}\left\vert E\left( x,t\right) T\left( x\right) e^{i\phi }\right\vert
^{2}\right\rangle \left\langle \sum_{q}\left\vert E\left( x,t\right)
\right\vert ^{2}\right\rangle .
\end{equation}%
Equation (1) shows that turbulence-free images cannot be reconstructed by
computational ghost imaging because the calculated light $\sum_{q}E\left(
x,t\right) $ is not affected by the same atmospheric turbulence as the real
light $\sum_{p}E\left( x,t\right) T\left( x\right) e^{i\phi }$.

To solve this problem, we duplicated the signal produced by the conventional
computational ghost imaging device into two identical copies. Thus, both
sets of data were affected by the same atmospheric turbulence. As a result,
the conditions for turbulence-free imaging were met. Then, turbulence-free
images were obtained by analysing both data sets based on the photon number
fluctuation autocorrelation. It was assumed that one image was generated for
every $m$ measurements. Thus, after $m\times n$ measurements, $n$ images
were produced. These $n$ images were duplicated (see the data processing
section in Fig. 1). The photon number fluctuation autocorrelation algorithm
was used to process these two data sets. The autocorrelation algorithm for
the photon number fluctuation is described below. The software first
calculated the average number of counts in a short time window $\overline{C}$%
. Two virtual logic circuits (post-neg identifiers) classified the number of
counts per window as positive or negative fluctuations based on $\overline{C}
$. Thus, we have

\begin{align}
\Delta C_{\alpha }^{(+)}& =\left\{
\begin{array}{c}
C_{\alpha }-\overline{C},if,C_{\alpha }>\overline{C} \\
0,otherwise%
\end{array}%
\right.   \notag \\
\Delta C_{\alpha }^{(-)}& =\left\{
\begin{array}{c}
C_{\alpha }-\overline{C},if,C_{\alpha }<\overline{C} \\
0,otherwise,%
\end{array}%
\right.
\end{align}%
where $\alpha =1$ to $n$ is used to label the $\alpha $th short time
window, $\Delta C$ is the photon number fluctuation. Then, we defined the following quantities for the statistical
correlation calculations:

\begin{align}
\left( \Delta C\Delta C^{\prime }\right) _{\alpha }^{\left( ++\right) }&
=\left\vert \Delta C_{\alpha }^{(+)}\Delta C_{\alpha }^{^{\prime
}(+)}\right\vert ,  \notag \\
\left( \Delta C\Delta C^{\prime }\right) _{\alpha }^{\left( --\right) }&
=\left\vert \Delta C_{\alpha }^{(-)}\Delta C_{\alpha }^{^{\prime
}(-)}\right\vert ,  \notag \\
\left( \Delta C\Delta C^{^{\prime }}\right) _{\alpha }^{\left( +-\right) }&
=\left\vert \left( \overline{C}-\Delta C_{\alpha }^{(+)}\right) \left(
\overline{C}-\Delta C_{\alpha }^{^{\prime }(-)}\right) \right\vert ,  \notag
\\
\left( \Delta C\Delta C^{^{\prime }}\right) _{\alpha }^{\left( -+\right) }&
=\left\vert \left( \overline{C}-\Delta C_{\alpha }^{(-)}\right) \left(
\overline{C}-\Delta C_{\alpha }^{^{\prime }(+)}\right) \right\vert .
\end{align}%
The statistical fluctuation-fluctuation correlation $\left\langle \Delta
C\Delta C^{^{\prime }}\right\rangle $ is then calculated from%
\begin{align}
& \left\langle \Delta C\Delta C^{^{\prime }}\right\rangle =\frac{1}{n}\left[
\sum_{\alpha =1}^{n}\left( \Delta C\Delta C^{^{\prime }}\right) _{\alpha
}^{\left( ++\right) }+\sum_{\alpha =1}^{n}\left( \Delta C\Delta C^{^{\prime
}}\right) _{\alpha }^{\left( --\right) }\right.   \notag \\
& +\left. \sum_{\alpha =1}^{n}\left( \Delta C\Delta C^{^{\prime }}\right)
_{\alpha }^{\left( +-\right) }+\sum_{\alpha =1}^{n}\left( \Delta C\Delta
C^{^{\prime }}\right) _{\alpha }^{\left( -+\right) }\right] .
\end{align}%
We have [28-30]
\begin{equation}
\left\langle \Delta C\Delta C^{^{\prime }}\right\rangle \propto \sum_{p\neq
q}\left\vert E_{p}^{^{\ast }}\left( x,t\right) E_{q}\left( x^{^{\prime
}},t\right) E_{q}\left( x^{^{\prime }},t\right) E_{p}^{^{\ast }}\left(
x,t\right) T^{2}\left( x\right) \right\vert ^{2}e^{i2\left( \phi +\phi
^{^{\prime }}\right) }
\end{equation}%
At $x=x^{^{\prime }}$, $e^{i2\left( \phi +\phi ^{^{\prime }}\right) }=1$,
and the cross-interference term reaches its turbulence-free constructive
maximum.

\section{Experiments and results}

The experimental setup was a conventional computational ghost imaging setup,
except for the data processing part. A semiconductor laser beam ($\lambda=$%
532 nm, 30 mW, Changchun New Industries Optoelectronics Technology Co., Ltd.
MGL-III-532) was used to illuminate a two-dimensional amplitude-only
ferroelectric liquid crystal SLM (FLC-SLM, Meadowlark Optics A512-450-850),
which had 512$\times$512 addressable 15$\mu m\times$15$\mu m$ pixels. Then,
the modulated light illuminated the object. Finally, the reflected light,
which carried information about the object, was collected by a bucket
detector. In this experiment, atmospheric turbulence was introduced by
adding conventional heating elements below certain sections of the optical
path. Heating the air causes temporal and spatial fluctuations in the index
of refraction, causing the traditional image of the object to jitter
randomly in the image plane and become distorted. The length of the heating
element was 0.4 m. The heating temperature was approximately 220$^{\circ}$C.
The photon number fluctuation autocorrelation algorithm was written in
MATLAB 2014a.
\begin{figure}[ptbh]
\centering\includegraphics[width=0.92\linewidth]{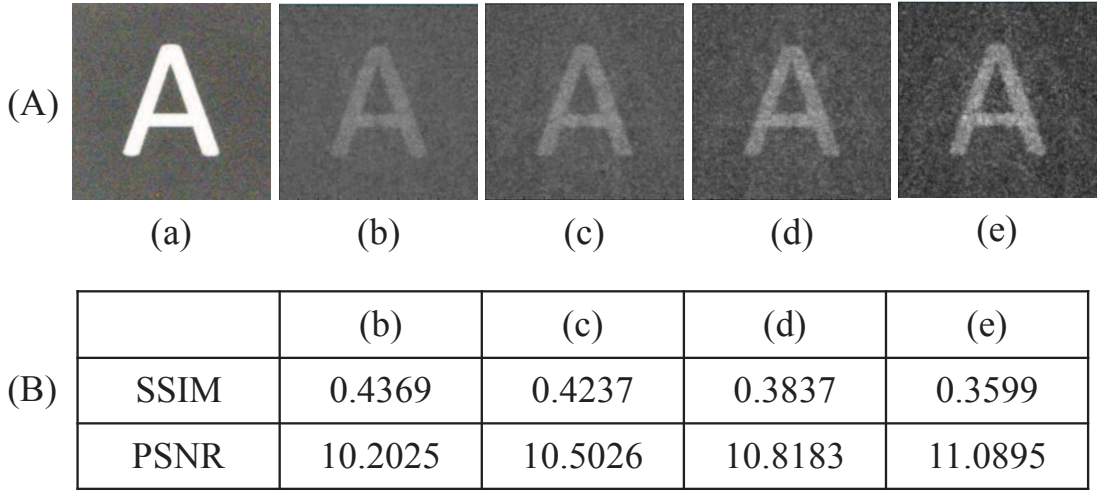}
\caption{(A) (a) The object and (b-e) the reconstructed images for different
measurements. (b) $m=500$, $n=20$, (c) $m=1000$, $n=10$, (d) $m=2000$, $n=5$%
, (e) $m=5000$, $n=2$. (B) The SSIM and PSNR values of the reconstructed
images.}
\end{figure}

We assumed that the computational ghost imaging device generated an image
every $m$ measurements. After $m\times n$ measurements, $n$ images were
produced. The first experimental results demonstrated that the images
reconstructed by our method were related to both $m$ and $n$. Fig. 2 clearly
shows that when $m\times n=10000$, the greater $m$ is, the brighter the
reconstructed image, but the lower the image quality. We used the structural
similarity (SSIM) index and the peak signal-to-noise ratio (PSNR) to
quantitatively measure the imaging quality of this method and conventional
computational ghost imaging [24].
\begin{figure}[ptbh]
\centering\includegraphics[width=0.92\linewidth]{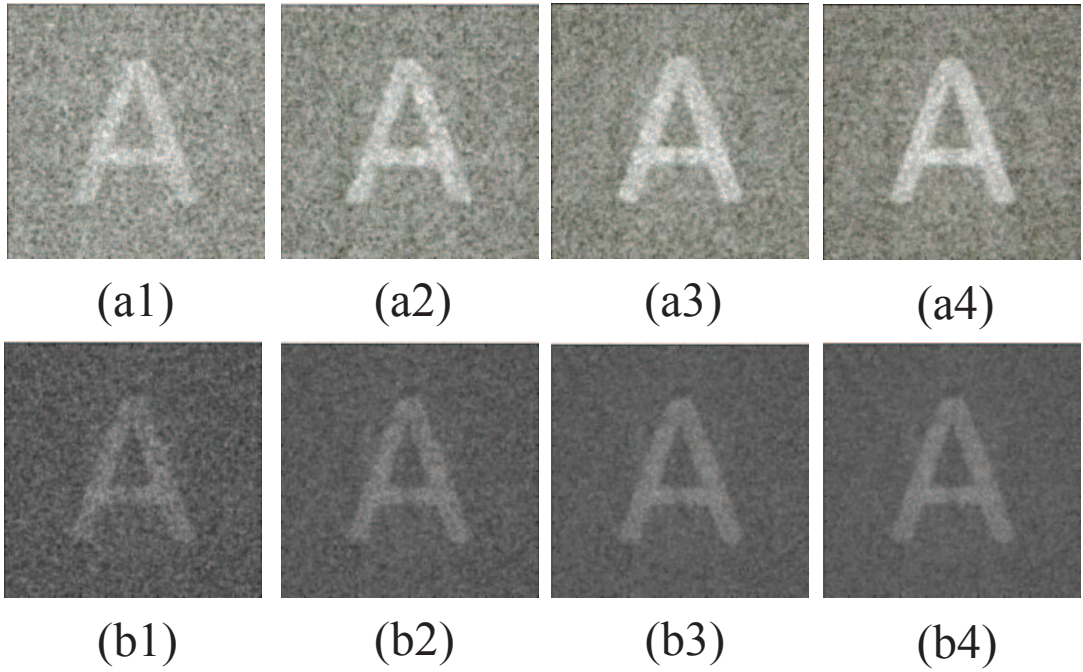}
\caption{Top row: the reconstructed images obtained by computational ghost
imaging with different measurements. (a1-a4) 1000, 4000, 7000, 10000. Bottom
row: the reconstructed images obtained by our method. (b1)$m=500, n=2$, (b1)$%
m=500, n=8$, (b1)$m=500, n=14$, (b1)$m=500, n=20$.}
\end{figure}
\begin{figure}[ptbh]
\centering\includegraphics[width=1\linewidth]{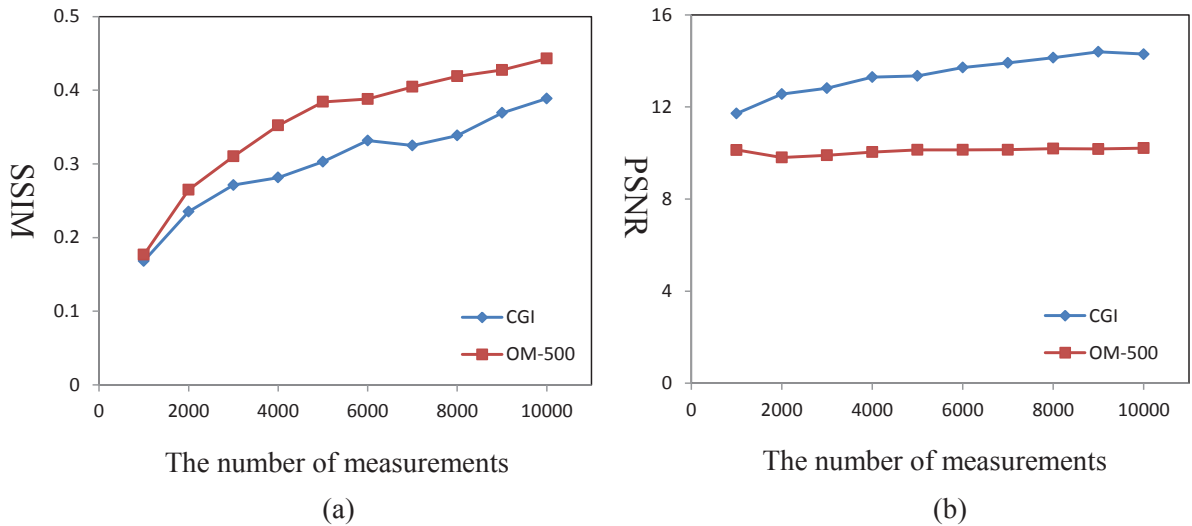}
\caption{The (a) SSIM and (b) PSNR curves of the images reconstructed by
computational ghost imaging and our method, respectively. CGI: computational
ghost imaging, OM-500: our method with $m=500$.}
\end{figure}
\begin{figure}[ptbh]
\centering\includegraphics[width=0.8\linewidth]{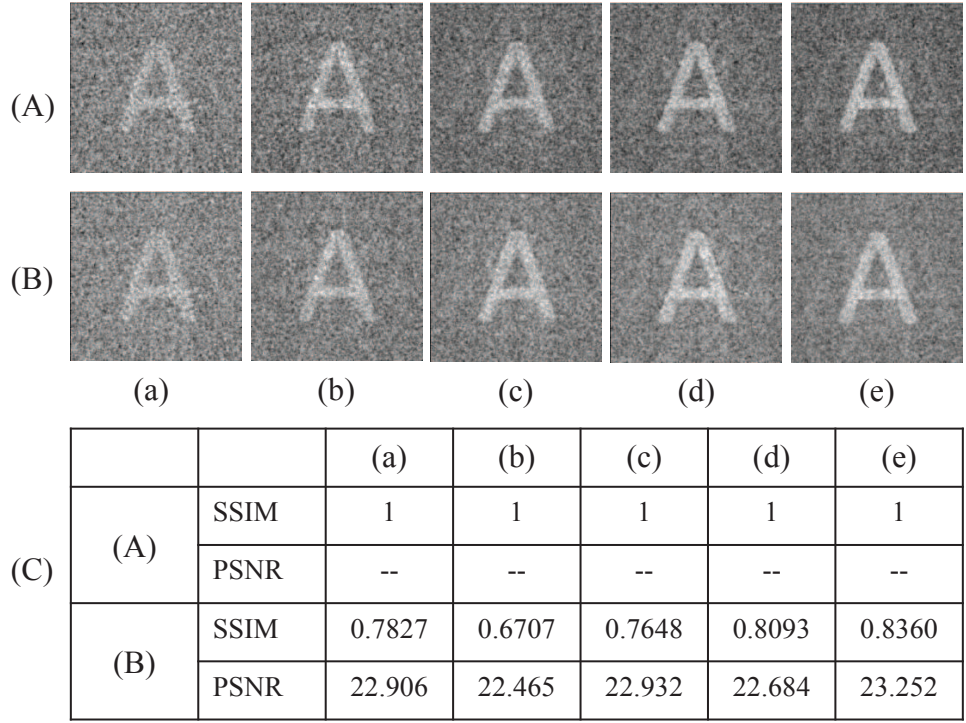}
\caption{(A) Computational ghost images without atmospheric turbulence. (B)
Distorted images affected by atmospheric turbulence. (a-e) Measurement times
of 1000, 2000, 3000, 4000 and 5000, respectively. (C) The SSIM and PSNR
values of the images corresponding to the above figures.}
\end{figure}
\begin{figure}[ptbh]
\centering\includegraphics[width=0.8\linewidth]{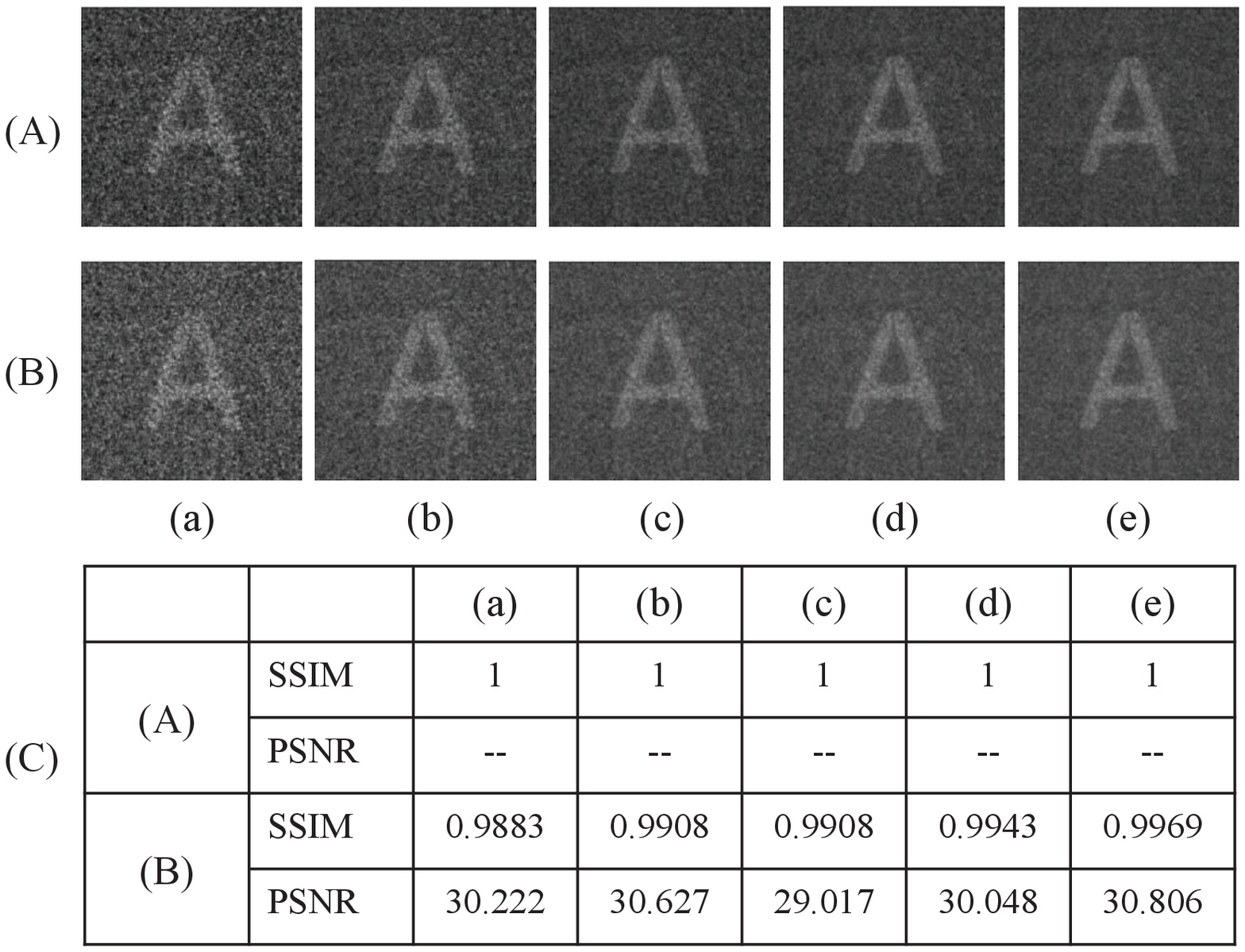}
\caption{(A) Images reconstructed by our method without atmospheric
turbulence. (B) Images reconstructed by our method with atmospheric
turbulence. (a-e) Measurement times of 1000, 2000, 3000, 4000 and 5000,
respectively, with $m=500$. (C) The SSIM and PSNR values of the images
corresponding to the above figures.}
\end{figure}

In the second experiment, we compared the quality of images reconstructed by
conventional computational ghost imaging and this method. Figure 3 shows
that the background noise of images reconstructed by this method was lower
than that of images reconstructed by conventional computational ghost
imaging. Moreover, Fig. 3 shows that the quality of the reconstructed images
improved as the measurement time increased. Fig. 4 shows that for the same
number of measurements, the quality of the images reconstructed by our
method was higher than that of the images reconstructed by conventional
computational ghost imaging (SSIM index), but the image brightness (PSNR
index) was lower than that of computational computational ghost imaging.

In the third experiment, we demonstrated that this method produced
turbulence-free images. The effect of turbulence on conventional
computational ghost imaging is shown in Fig. 5, while the results
reconstructed by our method are shown in Fig. 6. Fig. 5A shows a typical
result of computational ghost imaging in the absence of atmospheric
turbulence, while Fig. 5B shows the result of computational ghost imaging
with atmospheric turbulence. Figure 5C shows a quantitative comparison.
Here, we used the computational ghost images without atmospheric turbulence
as reference images. Similarly, Fig. 6A shows the result of our method
without atmospheric turbulence, while Fig. 6B shows the corresponding result
with atmospheric turbulence. Considering the fluctuation of light source, it
is clear that the images reconstructed by our method are turbulence-free
images.

\section{Conclusion}

These experiments demonstrate turbulence-free computational ghost imaging.
The problem that the calculated light differs from the actual light affected
by turbulence can be solved by measuring the photon number fluctuation
autocorrelation of the signal generated by a computational ghost imaging
device. Moreover, the quality of the images reconstructed by this method was
related not only to the total amount of data, but the number of images
generated by the computational ghost imaging device. When the sampling
amount is fixed, the fewer measurements per image produced by the
computational ghost imaging device, the better the overall result. We hope
that this method can provide a promising solution to overcome issues with
atmospheric turbulence in remote sensing and lidar applications.

This work was supported by the National Natural Science Foundation of China (11704221,11574178, 61675115); Taishan Scholar Foundation of Shandong Province (tsqn201812059).

Disclosures: The authors declare no constricts of interest.

Data Availability Statement:
Data underlying the results presented in this paper are not publicly available at this time but may
be obtained from the authors upon reasonable request.

\end{document}